\documentclass[prd,a4paper,aps,showpacs,twocolumn,floats]{revtex4}
\usepackage{epsfig}
\usepackage{amssymb}

\newcommand{\bk}{{\mathbf k}}

\newcommand{\bx}{{\mathbf x}}
\newcommand{\bz}{{\mathbf z}}

\newcommand{\HH}{{\cal H}}

\newcommand{\de}{\delta}
\newcommand{\De}{\Delta}
\newcommand{\ep}{\epsilon}

\newcommand{\La}{\Lambda}

\newcommand{\Om}{\Omega}

\newcommand{\ra}{\rightarrow}

\newcommand{\be}{\begin{equation}}
\newcommand{\ee}{\end{equation}}
\newcommand{\bea}{\begin{eqnarray}}
\newcommand{\eea}{\end{eqnarray}}
\newcommand{\bean}{\begin{eqnarray*}}
\newcommand{\eean}{\end{eqnarray*}}
\newcommand{\dd}{\partial}
\newcommand{\mr}{\mathrm}

\newcommand{\eg}{{\em e.g. }}

\def\lsim{\, \raise 0.4ex\hbox{$<$}\kern -0.8em\lower 0.62
ex\hbox{$\sim$} \,}
\def\gsim{\raise 0.4ex\hbox{$>$}\kern -0.8em\lower 0.62
ex\hbox{$\sim$}}

\begin{document}

\title{
CMB anisotropies from acausal scaling seeds
}
\author{Sandro Scodeller}
\email{sandro.scodeller@astro.uio.no}
\affiliation{Institute of Theoretical Astrophysics, University of 
Oslo,\\PO Box 1029 Blindern, N-0315 Oslo, Norway}

\author{Martin Kunz}
\email{m.kunz@sussex.ac.uk}
\affiliation{Astronomy Centre, University of Sussex, Brighton BN1 9QH, UK}
\affiliation{D\'epartement de Physique Th\'eorique, Universit\'e
de Gen\`eve, 24 quai Ernest Ansermet, CH--1211 Gen\`eve 4,
Switzerland}

\author{Ruth Durrer}
\email{ruth.durrer@unige.ch}
\affiliation{D\'epartement de
Physique Th\'eorique, Universit\'e de
Gen\`eve, 24 quai Ernest Ansermet, CH--1211 Gen\`eve 4, Switzerland}

\date{\today}

\begin{abstract}
We investigate models where structure formation is initiated by scaling seeds:
We consider rapidly expanding relativistic shells of energy and show that 
they can fit current CMB and large scale structure data if they expand with
super-luminal velocities. These acausally expanding shells provide a viable 
alternative to inflation for cosmological structure formation with the same 
minimal number of parameters to characterize the initial fluctuations. 
Causally expanding shells alone cannot fit present data.  Hybrid 
models where causal shells and inflation are mixed also provide good fits.
\end{abstract}


\maketitle
\section{Introduction}
\label{intro}
Inflationary models provide an excellent fit to all fluctuation data, the cosmic
microwave background (CMB) anisotropies and polarization, as well as large
scale structure data from galaxy catalogs. However, most current inflationary 
scenarios are simple toy models which are not well motivated from high energy 
physics. It is therefore not only justified, but important to 
study other ways to generate initial fluctuations. 

In the past, especially models where topological defects act as seeds for 
fluctuations of the matter--radiation fluid have been studied. For simple 
global topological defects and for
cosmic strings from the Abelian Higgs model, it has been found that they cannot
reproduce the inflationary peak structure predicted by inflation~\cite{DSG,CHM}.
Comparison with present observations shows that topological defects can
contribute at most about 10\% to the CMB temperature fluctuations
on large scales~\cite{martin}.

Since these competing models for structure formation have been ruled out,
the general point of view in the field seems to be 
that only inflation can lead to a
coherent series of acoustic peaks. However, this is not correct: Neil Turok
and others have shown~\cite{turok,hsw} that a scaling seed model, where the 
seeds consist of a
stochastic distribution of rapidly expanding shells of energy with a velocity
close to the speed of light also leads to an acoustic peak structure like
inflation. This ``Turok model'' was very promising in fitting CMB
data back in 2001~\cite{DKMseeds}. However, there are arguments~\cite{SpeZa}
that causal scaling seeds are not able to produce the a first peak in the 
E-polarization spectrum at $\ell \simeq 100$, which is due to polarization 
at the last scattering surface where this scale was super-horizon.
As we show here, the Turok model with shells which expand with sub-luminal 
velocities (causal shells) indeed cannot fit the T-E correlation spectrum in  
present CMB data. 
However, this can be evaded if we allow for super-luminal expansion of the 
shells (acausal shells). As we shall see, acausal shells can
generate CMB anisotropies and polarization as well as large scale structure 
which are in good agreement with present data. They
fit the WMAP 3 year~\cite{w3dat} 
and ACBAR '08~\cite{ACBAR08} data as well as a simple inflationary model 
with the same number of parameters. Actually our simple model has three 
parameters to describe the seed perturbations, the amplitude and two 
velocities, however, as we shall see, two of them are strongly correlated, 
so that the parameter space is effectively two dimensional like in the
simplest inflationary models (without gravitational waves and without running).

Super-luminal velocities are usually considered unphysical as they generically
lead to signals which can propagate along a closed loop, see e.g.~\cite{causal}.
However, in the cosmological situation where we have a preferred Lorentz 
frame, the cosmological time, this conclusion can be avoided~\cite{MuBa} 
since boost symmetry is broken. 

In Ref.~\cite{causal} it is shown that, if we 
treat all observers equally and demand that they can only send signals 
forward in time with respect to their proper time, super-luminal motion 
leads to closed signal curves and all the difficulties that come with it
(information can be sent from the future into the past, the concept of 
entropy becomes ill-defined etc.). This is 
equivalent to asking that the propagation equations for a field has to be 
solved using the retarded Green function w.r.t. all reference frames.
In Minkowski space this seems the only viable possibility since all frames 
are equivalent. 

However, as has been argued in Ref.~\cite{MuBa}, on a
background which breaks boost symmetry, like the Friedmann-Lema\^\i tre
solutions, one does have a preferred reference frame (cosmological time)
and not all reference frames are equivalent. One can therefore prescribe
the initial conditions for the propagation equation of a field such that
the retarded Green function with respect to the cosmological frame is 
always used. Then, by construction, all signals travel forward in 
cosmological time and no closed signal curves can emerge. Let us now
consider a mode (like e.g. our exploding shells) which propagates faster 
than the speed of light, say with speed $v>1$. An observer moving with 
velocity $v_o$ so that $1>v_o>1/v$ with respect to the cosmological frame, 
then sees signals propagating from the future into the past (in certain 
directions, see Ref.~\cite{causal} for details). This
is very unusual to say the least, but cannot be excluded by experiment.
Especially 
if the particles which make up the exploding shells interact 
with standard model fields only gravitationally, one cannot see this
effect on small scales, i.e. in laboratory experiments. However, as we shall
argue in this paper, on cosmologically large scales these exploding shells 
can become important.

In the remainder of this paper we show that super-luminally 
expanding shells can fit present CMB data as well as inflationary models.
In the next section we define and discuss our seed model. In Section III 
we present the results for the cosmological parameters as well as the 
primordial parameters for the pure seed model and the hybrid model. In the 
last section we draw some conclusions.

In this paper conformal time is denoted by $t$ so that the metric is
$$ ds^2 = a^2(t)\left(-dt^2 + \de_{ij}dx^idx^j\right)~.$$
We set the spatial curvature to zero. Spacetime indices are lower case Greek 
letters and 3D spatial indices are lower case Latin letters.
The conformal Hubble parameter is $\HH=\dot a/a = aH$, where $H$ denotes the 
physical Hubble parameter and a dot is a derivative w.r.t. conformal time $t$.

\section{The models}
\label{model}
We consider an inhomogeneous uncorrelated distribution of spherical
expanding shells. The energy momentum tensor of uncorrelated spherical 
shells is purely scalar and we can parameterize it in the following way
\bea
T^0_0 &=& -\frac{M^2}{a^2}f_\rho ~,\\
T^i_j &=& \frac{M^2}{a^2}\left[f_p\de^i_j + \left(\dd_i\dd_j -
      \frac{1}{3}\de^i_j\De\right)f_\pi \right] ~,\\
T^0_i &=& \frac{M^2}{a^2}\dd_if_v ~.\eea  
The energy density plus three times the pressure as well as the energy flux of the shells are
posited to be
\be
f_\rho(\bx,t)+3f_p(\bx,t) = \sum_n 
\frac{\de(|\bx-\bz_n|-v_1t)}{4\pi \HH t^{3/2}|\bx-\bz_n|^2}~, \ee \be \label{eq:frx}
f_v(\bx,t) = -\sum_n 
  \frac{3E(t)\theta(v_2t-|\bx-\bz_n|)}{4\pi v_2^2|\bx-\bz_n|t^{3/2}}~.
\ee
Here the positions $\bz_n$ are the centers of the exploding shells which are at 
random, uncorrelated positions.
The function $\theta$ is the Heaviside function,
$$
\theta(y) =  \left\{\begin{array}{ll}1 & \mbox{if } y>0\\
0 & \mbox{else,} \end{array} \right.
$$
and the function $E(t)$ is given by
 $$ E(t)= \frac{4-2(\HH t)^{-1}}{3+12\HH t}~.$$
To simplify the analysis we use infinitely thin shells for which the inner and 
outer radii coincide and expand with the same velocity. (This corresponds 
to the limit $B\ra C$ in the original Turok model~\cite{turok}.)  
The above form of the energy momentum tensor ensures that the perturbations are
of purely scalar nature. The two remaining functions in the
parameterization of $T_\mu^\nu$
are determined by
energy and momentum conservation.  The choice of 
$E(t)$, together with the factor $1/v_2^2$, assures that also $f_\pi$ has
compact support, $f_\pi(\bx,t)=0$ if $|\bx-\bz_n|>vt$, where $v=\max(v_1,v_2)$.
Then $f_\pi$ has a white noise
spectrum on large scales, $ktv<1$. We assume that the centers $\bz_n$
of the shells are uncorrelated and have a fixed comoving space density.
Up to an irrelevant phase coming from the position of the shell
center, the Fourier transforms of the source functions from one shell
are given by
\be
 (f_\rho+3f_p)(\bk,t) = \frac{1}{\HH t^{3/2}}
   \frac{\sin(v_1kt)}{v_1kt} ~, \label{e:theta}\ee   \be
 f_v(\bk,t) = \frac{3E(t)}{v_2^2k^2t^{3/2}}
  \left(\cos(v_2kt)-\frac{\sin(v_2kt)}{v_2kt}\right)  \label{e:fv}\, .
\ee
As the different shells are uncorrelated, their contributions can
 be added with random phases. In the limit of many uncorrelated
shells, the spectra of $f_\rho+3f_p$ 
and $f_v$ are then simply the squares of the above functions,
e.g.,
\be
\langle f_v(\bk,t)f_v^*(\bk',t)\rangle = (2\pi)^3\de(\bk-\bk')P_v(k,t) 
\nonumber 
\ee
with
\bea
P_v(k,t) &=& A^2\frac{9E^2(t)}{v_2^4k^4t^{3}} \left(\cos(v_2kt)-
  \frac{\sin(v_2kt)}{v_2kt}\right)^2  \label{e:Pv} ,\\
 &=& A^2| f_v(k,t)|^2 .
\eea
The constant pre-factor $A$ determines the number of shells per Hubble volume.

 In Fourier space, energy and momentum conservation require
\bea
\dot f_\rho + k^2f_v +\HH(f_\rho+3f_p) =0\,, && \\
\dot f_v + 2\HH f_v -f_p +\frac{2}{3}k^2f_\pi =0\,.  \label{e:fpi}&& 
\eea
Integrating the first equation one finds \eg during the radiation
dominated era when $\HH t=1$, so that $E(t)= 2/15$,
\bea
\hspace*{-1.8cm}f_\rho &=& \frac{4}{3\sqrt{t}}\Big\{\Big[\cos(v_1kt) 
+\frac{\sin(v_1kt)}{2v_1kt}   \nonumber \\  && \quad\qquad
 +\sqrt{2\pi v_1kt}S\left(\sqrt{\frac{2v_1kt}{\pi}}\right) \Big]
 \nonumber \\  && \quad\qquad
+\frac{1}{5v_2^2}\Big[\cos(v_2kt)-\frac{\sin(v_2kt)}{v_2kt} +
\nonumber \\  && \quad\qquad  \label{e:frho}
  \sqrt{2\pi v_2kt}
    S\left(\sqrt{\frac{2v_2kt}{\pi}}\right) \Big]\Big\} ~.
\eea
\bea
f_\pi &=& \frac{1}{k^2\sqrt{t}}\Big\{\frac{3}{10}
  \Big[ 2\frac{\sin(v_2kt)}{v_2kt} +\frac{\cos(v_2kt)}{(v_2kt)^2} 
  -\frac{\sin(v_2kt)}{(v_2kt)^3}\Big]  \nonumber \\  && \qquad
-\frac{2}{15v_2^2}\Big[\cos(v_2kt)- \frac{\sin(v_2kt)}{v_2kt} +   \nonumber \\ 
   && \qquad \sqrt{2\pi v_2kt}
  S\left(\sqrt{\frac{2v_2kt}{\pi}}\right) \Big]  \nonumber \\  && \qquad
-\frac{2}{3}\Big[\cos(v_1kt)-\frac{1}{4}\frac{\sin(v_1kt)}{v_1kt} + 
 \nonumber \\  && \qquad  \sqrt{2\pi v_1 kt} 
S\left(\sqrt{\frac{2v_1kt}{\pi}}\right) \Big] \Big\}~.
\eea
Here $S(x)$ denotes the sine Fresnel integral as defined 
in~\cite{AS}. 
A similar result is obtained during the matter era. Even though this 
is not directly evident, a series expansion shows that $f_\pi$ is white 
noise for small arguments, $kt\ll 1$, as it should be for the Fourier
transform of a function with compact support.
The source function $f_p$ is easily determined with the help of
Eqs.~(\ref{e:theta}) and (\ref{e:frho}).
The metric perturbations due to the shells are given by the seed
Bardeen potentials~\cite{d90} 
\bea
k^2\Phi_s &=& \ep(f_\rho + 3\HH f_v) \,,\\
\Psi_s &=& -\Phi_s - 2\ep f_\pi \, ,\\
\mbox{where } \quad \ep & =& 4\pi GM^2A \ll 1
\eea
determines the overall amplitude.
The matter Bardeen potentials on the other hand are given by the
matter density perturbations and anisotropic stresses,
\bea
k^2\Phi_m &=& 4\pi Ga^2\rho D \,,\\
k^2(\Phi_m + \Psi_m) &=& 8\pi Ga^2p\Pi~,
\eea
where $\Pi$ denotes the anisotropic stress of the cosmic fluid and $D$
is a gauge invariant density perturbation variable. Care is required
when relating $D$ to the density fluctuation in longitudinal gauge
since then a term proportional to the total Bardeen potential
$\Phi=\Phi_s+\Phi_m$ enters the equation. More details can be found 
in Refs.~\cite{DKMseeds,master}.
The total Bardeen potentials,
\be \Phi=\Phi_s+\Phi_m ~, \mbox{ and } \quad
 \Psi=\Psi_s+\Psi_m
\ee
then enter the usual evolution equation for cosmic matter and radiation. 

We shall now show that even though current CMB data cannot be fitted with expanding 
shells as long as we require causality, when allowing for super-luminal expansion
we can obtain excellent fits which rival the fits from inflationary models
to all data.
We shall also comment on mixed models. 

\begin{figure}
\centerline{\epsfig{figure=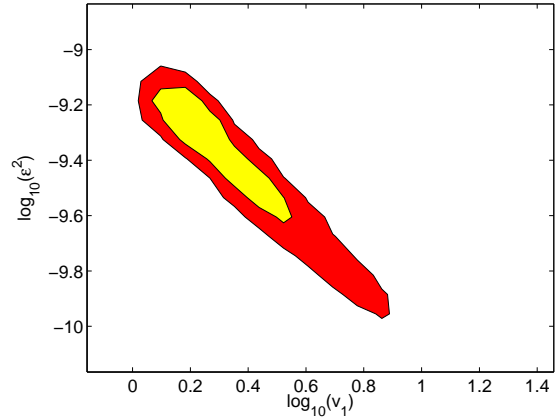,width=8cm}}
\caption{\label{f:v1amp} A 2D likelihood plot showing the
degeneracy between the velocity $v_1$ and
the amplitude $\epsilon^2$. The yellow colored area encloses 68\% and the
larger red colored area 95\% of the likelihood volume (marginalized
over all other parameters).}
\end{figure}

The seed perturbations from expanding shells are determined by the
velocities $v_1$ and $v_2$ and an amplitude $\ep$ which is proportional to
the number density of shells. However, the amplitude needed to obtain a 
good fit is tightly correlated with $v_1$, as we shall discuss below
and as is shown in Fig.~\ref{f:v1amp}.
Once the best fit value of $v_1$ is determined, the amplitude is effectively 
fixed by 
$$\ep^2=9.4\times10^{-10}/v_1~.$$ 
Therefore, expanding shells can be regarded as models requiring 
effectively two parameters for the initial fluctuations, like scalar 
inflationary perturbations.

For the mixed models, we add scalar fluctuations from inflation which are 
characterized by the amplitude $A_s$ and the scalar spectral index $n_s$,
$\langle|\Phi_m(t_\mr{in},k)|^2k^3\rangle \simeq 
\langle|\Psi_m(t_\mr{in},k)|^2k^3\rangle = A_s(k/k_0)^{(n_s-1)}$, where $k_0
 =0.002$Mpc$^{-1}$ is the pivot scale. We assume the inflationary 
perturbations to be uncorrelated with the seeds (expanding shells).

\section{Results}
\label{result}

We start by investigating models where perturbations are generated purely 
from the expanding shells. 
\begin{figure}
\centerline{\epsfig{figure=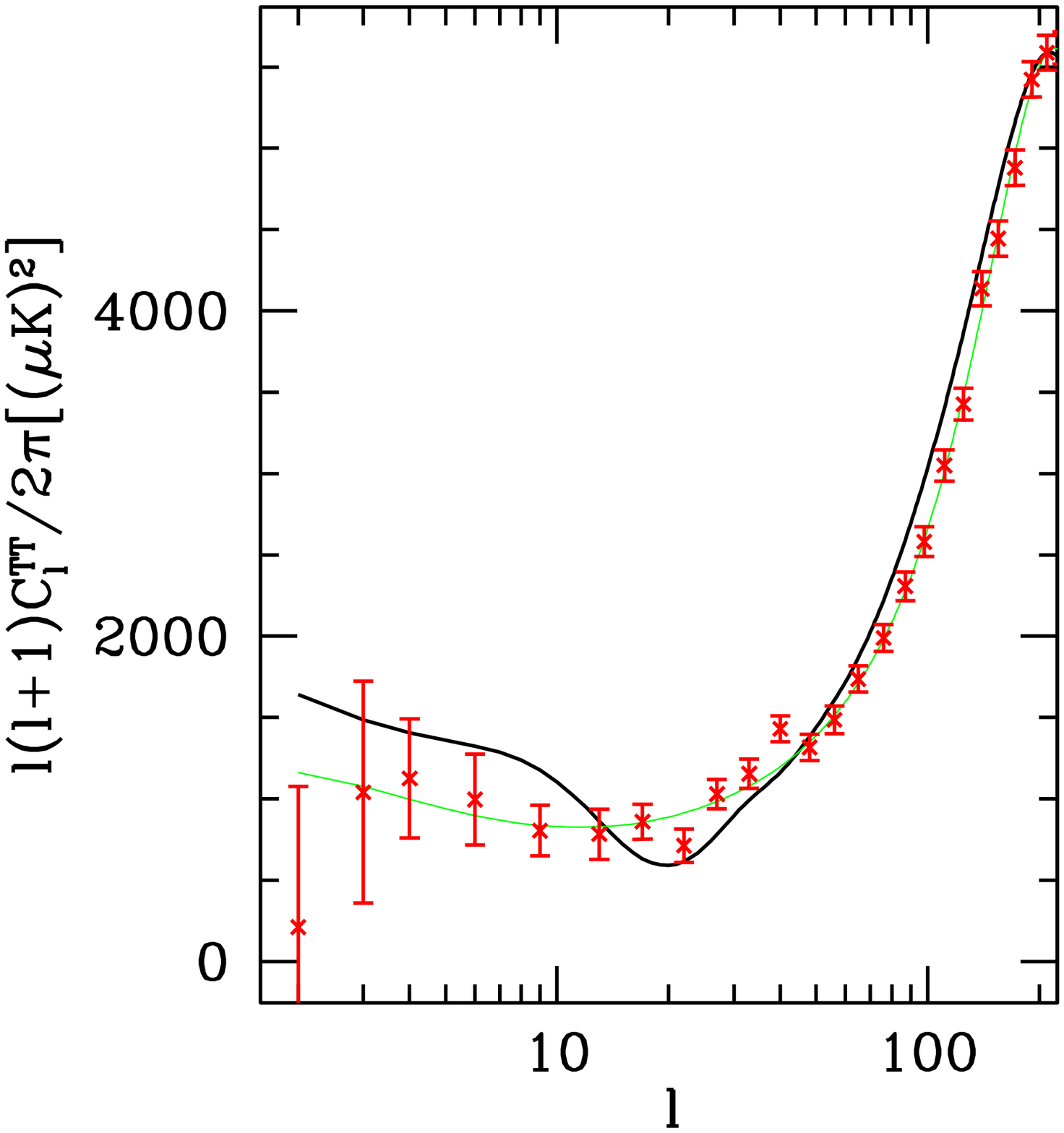,width=6.5cm}}
\centerline{\epsfig{figure=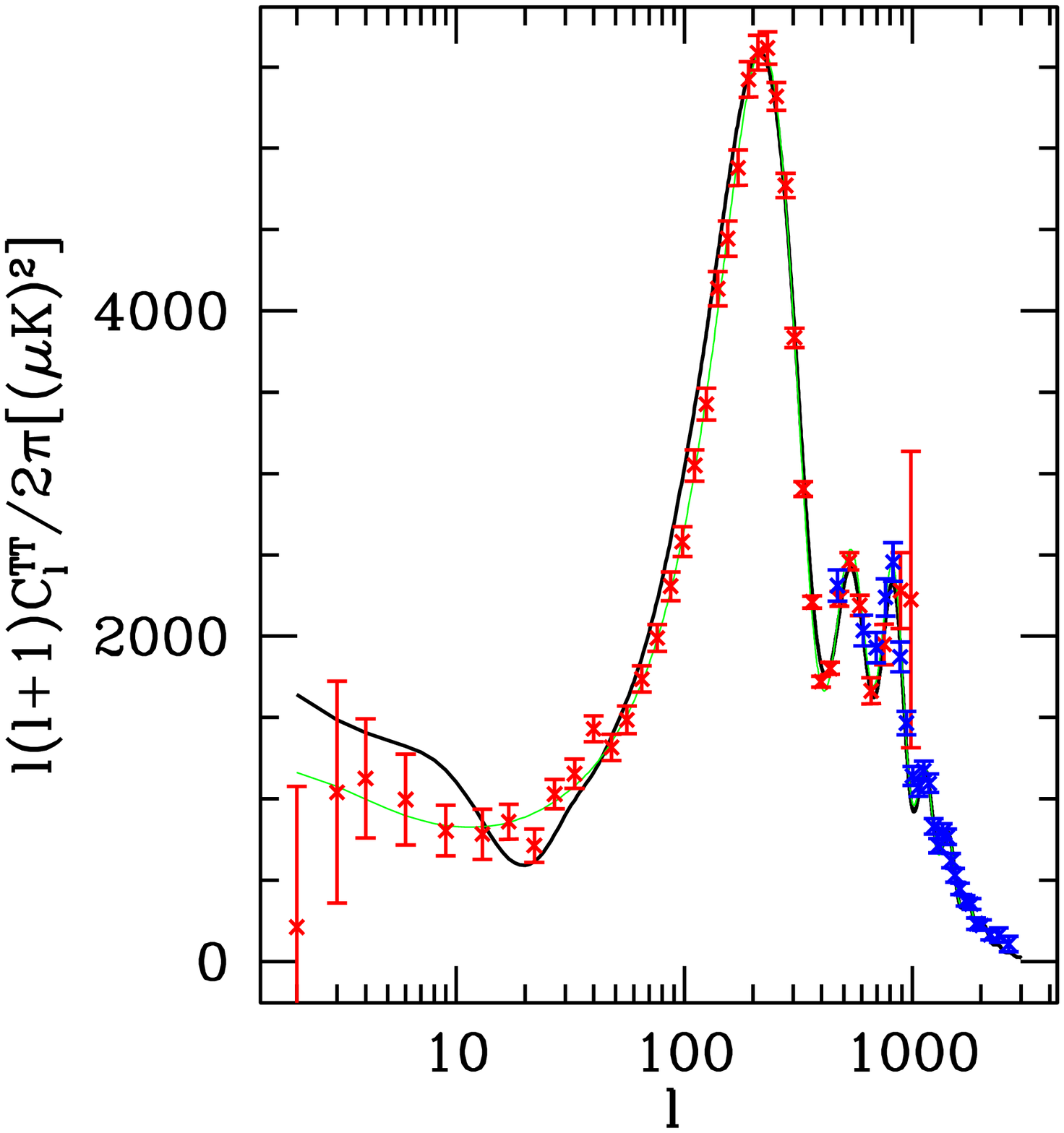,width=6.5cm}}
\caption{\label{f:TTcausal} The best fit CMB anisotropies from a
{\bf causal} model of expanding shells is shown (fat black line) and 
compared with the data from
  WMAP and ACBAR, and to the best fit $\La$CDM model (thin green
  line). The top panel shows the rise to the first peak, $\ell\leq 200$ 
which can not be fitted satisfactorily by this model. 
  The bottom panel shows the spectrum up to $\ell = 2500$. The 
secondary peaks are well fitted. The parameter values for  the best
fit causal shell model are given in the text.
}
\end{figure}
A first interesting result is that this model cannot provide a good fit to 
CMB data
if we constrain $v_1\le 1$ and $v_2\le 1$. We can obtain reasonable, but not 
sufficiently good fits for the temperature anisotropy, see 
Fig.~\ref{f:TTcausal}, and we cannot fit the polarization data. This is seen 
especially well when comparing the model with the high quality TE 
correlation data from WMAP~\cite{w3dat}, see Fig.~\ref{f:TEcausal}. 
\begin{figure}
\centerline{\epsfig{figure=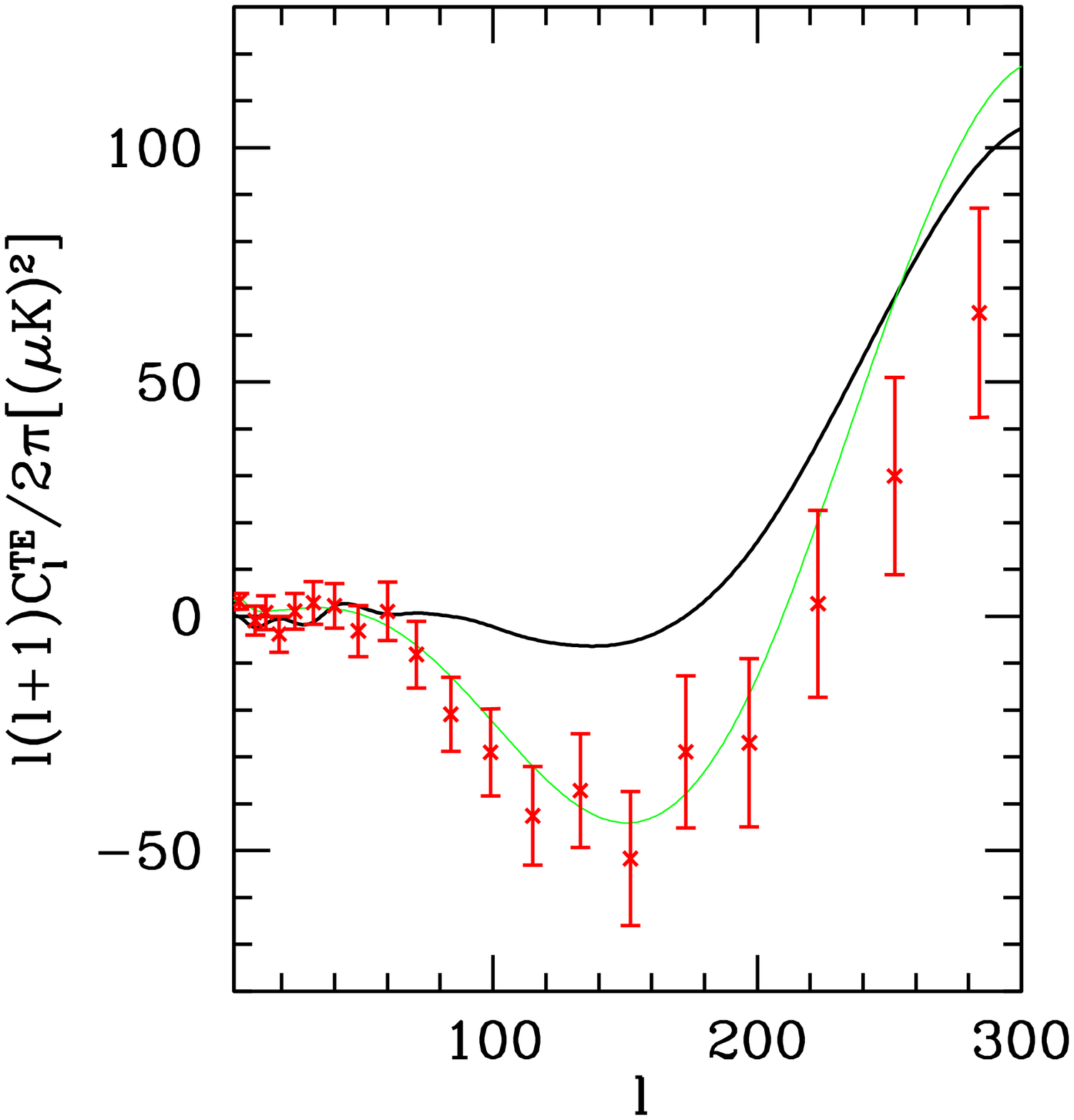,width=6.5cm}}
\centerline{\epsfig{figure=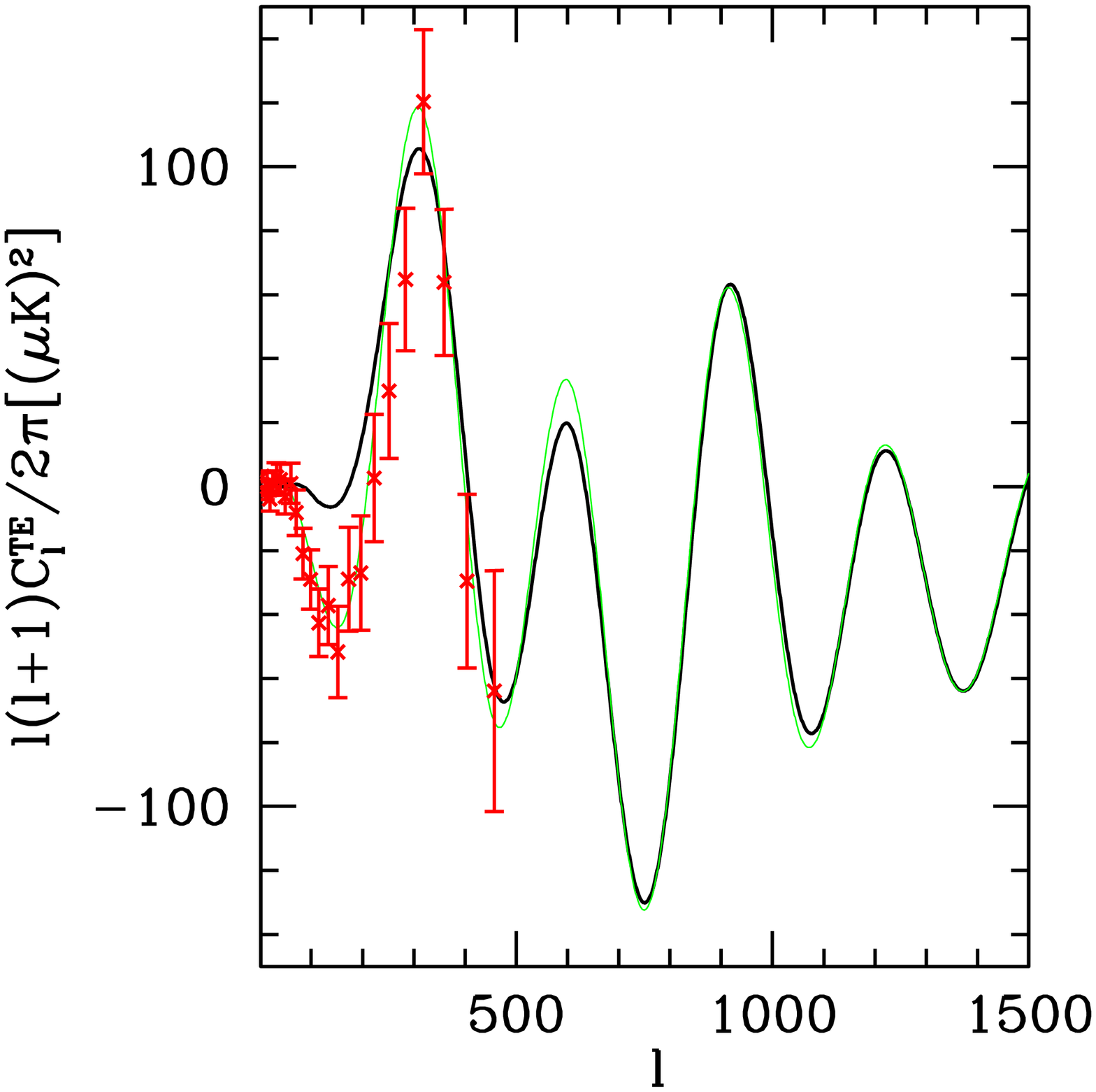,width=6.5cm}}
\caption{\label{f:TEcausal} The best fit T-E correlation spectrum from a
  {\bf causal} model  of expanding shells (fat black line) is compared 
with the data from WMAP and with a standard $\La$CDM model (thin green line). 
The top panel shows 
the first acausal anti-correlation peak, at $\ell\simeq 150$ which is absent in 
the causal model, while the bottom panel shows the spectrum up to 
$\ell = 1500$. The secondary peaks are very similar to the inflationary 
case. The parameter values are the same as for Fig.~\ref{f:TTcausal}.}
\end{figure}

\begin{figure}
\centerline{\epsfig{figure=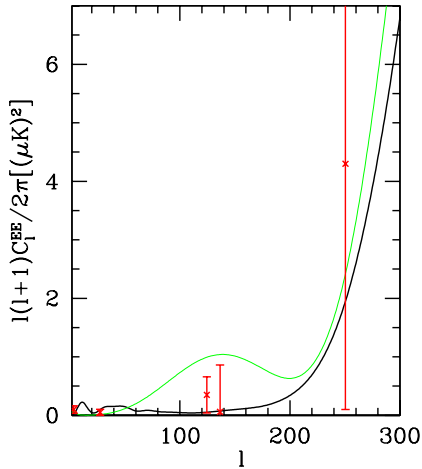,width=6.5cm}}
\caption{\label{f:EEcausal} The best fit EE power spectrum from a
  {\bf causal} model of expanding shells (fat black line) is compared with 
the data from WMAP and with a standard $\La$CDM model (thin green line). 
Only the region 
of the first acausal peak, $\ell\leq 300$ is shown, where the causal
model differs from inflation. The secondary peaks are very similar to 
the inflationary case. The parameter values are the same as for 
Fig.~\ref{f:TTcausal}.}
\end{figure}

Also the pure polarization
spectrum differs from the inflationary polarization by the absence 
of the first, acausal peak at $\ell\simeq 130$, see Fig.~\ref{f:EEcausal}.
However, the current observations of the EE spectrum are not sufficiently 
accurate on large scales to rule out the absence of a peak at $\ell \simeq 130$.

We have used the code CMBEASY~\cite{cmbeasy} and its Monte Carlo Markov Chain
(MCMC) analysis tool~\cite{analysit} to determine the best fit
cosmological parameters for a spatially flat cosmology with photons,
massless neutrinos, cold dark matter and a cosmological constant.
For the fitting procedure we used the 3 year WMAP data~\cite{w3dat,page}, 
the Boomerang 2003 data~\cite{Boom03}, the CBI~\cite{cbi} and
the old ACBAR data~\cite{acbar06}, as well as the Sloan Digital Sky Survey 
(SDSS) power spectrum for luminous red galaxies~\cite{lrg}, which is 
supposed to be still in the linear regime.  In the  figures the best 
fit solution for the CMB anisotropies are compared with the 3 year WMAP
data~\cite{w3dat,page} and  with the recent ACBAR
results~\cite{ACBAR08}. 

The maximum of the likelihood for the causal models has quite
a complicated structure, with several peaks close together.
The best-fitting model which we could find has the following
parameter values: 
a Hubble parameter $H_0 =100h$km/s/Mpc where $h=0.686$, a matter density 
parameter $\Om_mh^2=0.137$, a baryon density parameter $\Om_bh^2=0.0220$, 
an optical depth $\tau =0.36$, shell velocities $v_1=0.77$, 
$v_2=1.0$ and the amplitude $10^{10}\ep^2 =26.0$. Most cosmological parameters 
are similar to their values for inflationary perturbations. 
The best fit parameters for a simple inflationary model fitted to the 
same data are $h=0.713$, $\Om_mh^2=0.133$, $\Om_bh^2=0.0223$, 
$\tau =0.08$, $n_s=0.956$ and $A_s=2.3\times 10^{-9}$.

The best fit optical depth for the causal seed model is larger
than in the inflationary models. In order to generate T-E correlations on 
large scales, the optical depth tends to increase. The best fit
velocity $v_2$ is at the upper limit of the prior and would prefer to
exceed the causality limit. Our MCMC chains had severe difficulties to 
converge for this model, which is partly due to the fact that the
best fit lies at the boundary of the priors for several parameters, which
seems to cut a connected acausal best-fit region up into several unconnected
causal ones. We can 
therefore not say that a velocity $v_1<1$ is significantly preferred since
there is a second maximum with $v_1\approx 1$ and $v_2\approx0.7$ which
seems to have $\Delta\chi^2=20$ with respect to the best-fit models, but
which is completely disconnected from the first maximum
 so that no chains have managed to sample both.
We are also reluctant to quote 1-sigma errors, first of all since the MCMC
chains did not converge well, and secondly since the model is 
not a good fit and hence error-bars are not useful.

Only if we allow for super-luminal 
expansion of the shells can we obtain a good fit to present data. 
The best fit cosmological parameters for super-luminally expanding shells
obtained using the same data are surprisingly close to those for an 
inflationary $\La$CDM model: 
We find $\Om_mh^2 = 0.134$, $\Om_bh^2 =0.0232 $, $h=0.745$, $\tau =0.11$. 
The best fit model parameters are $v_1=1.65$ and
$v_2=5.66$. The amplitude is inversely
proportional to $\sqrt{v_1}$ and the CMB normalization requires
$\ep^2v_1=9.4\times10^{-10}$, see Fig.~\ref{f:v1amp}. 
For the above value of $v_1$ we therefore infer $\ep^2 =5.7\times10^{-10}$. 

\begin{figure}
\centerline{\epsfig{figure=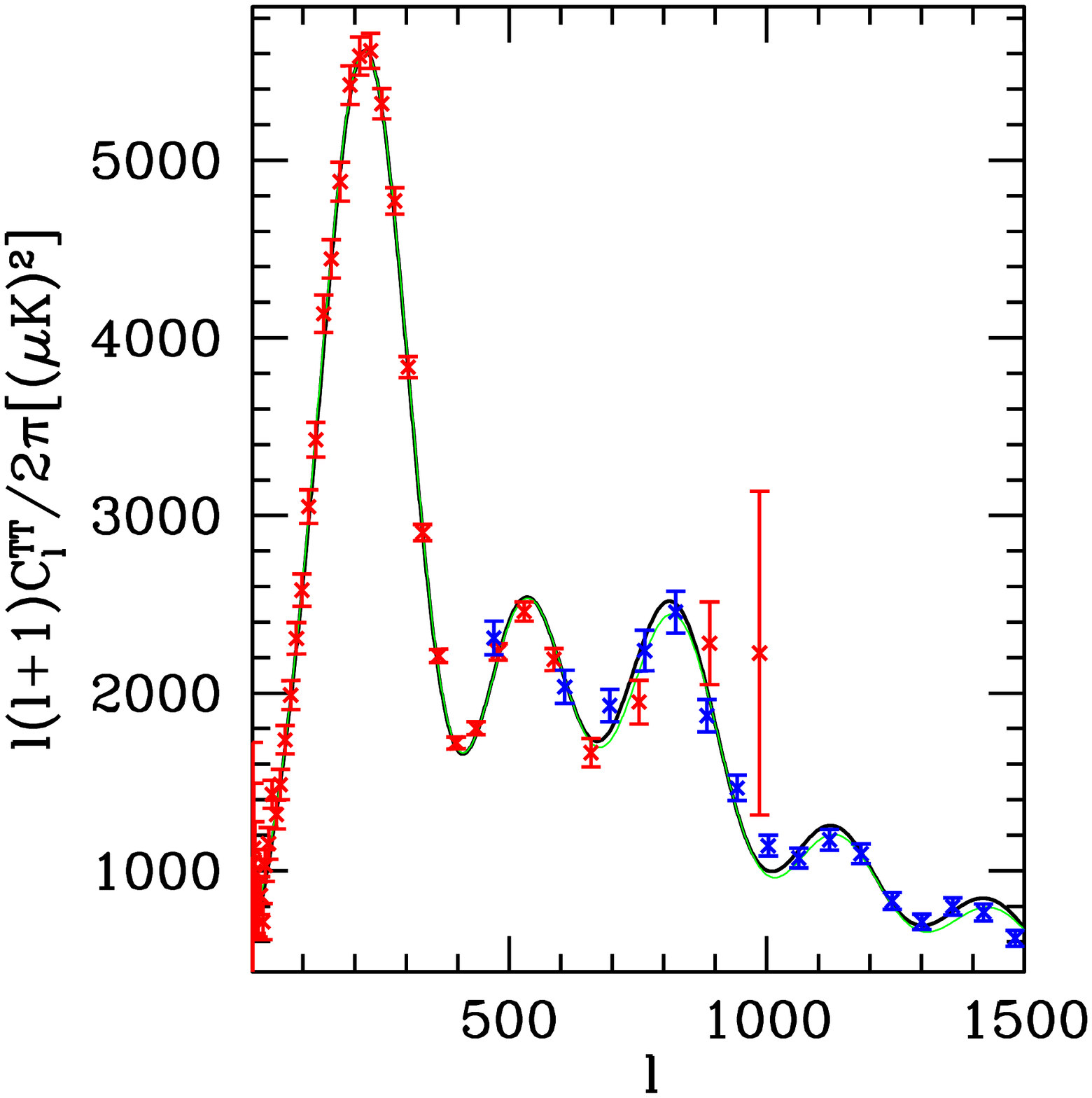,width=6.5cm}}
\centerline{\epsfig{figure=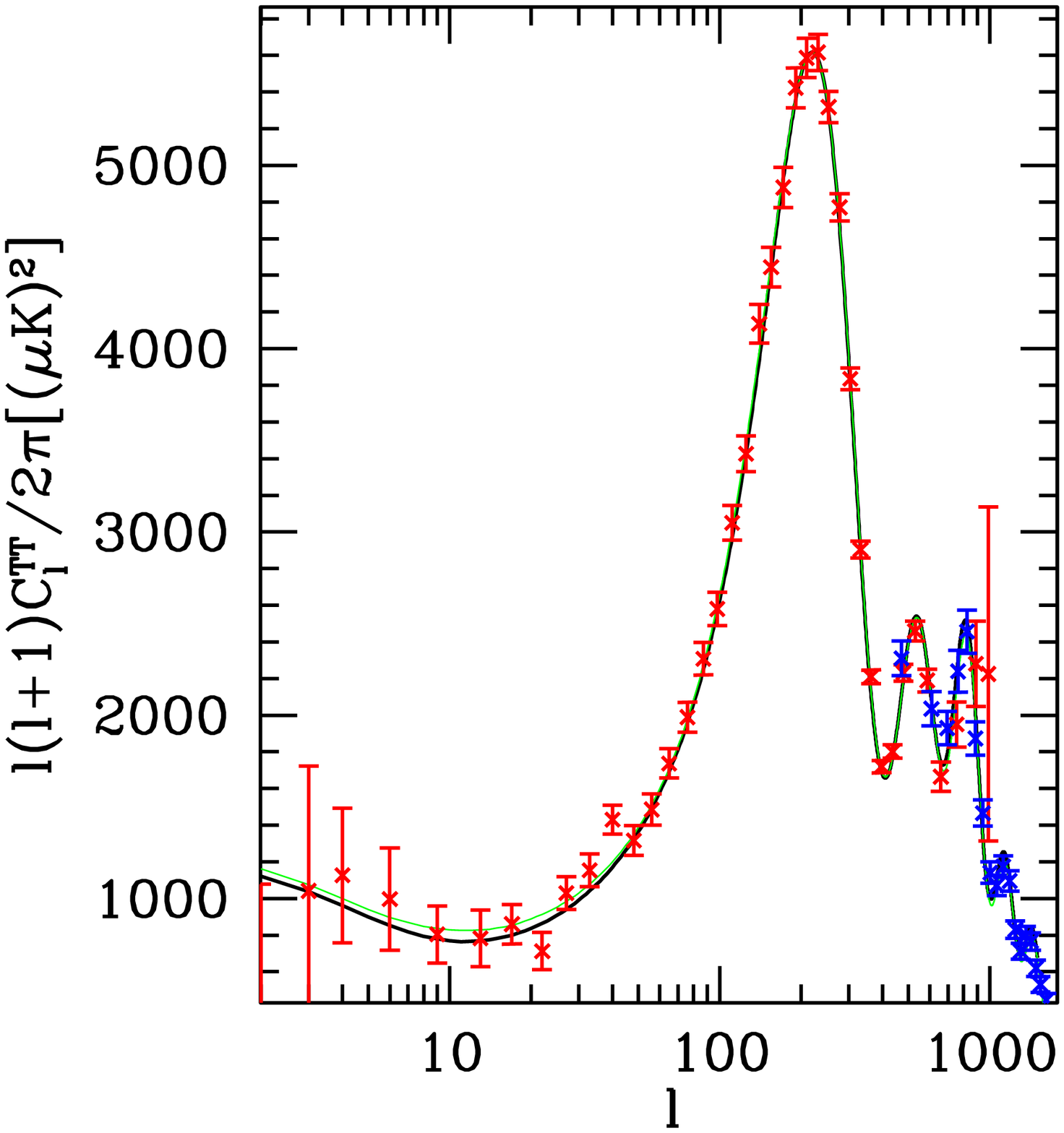,width=6.5cm}}
\caption{\label{f:pure1} The best fit CMB anisotropies from an {\bf acausal}  
seed model are shown (fat black line) and compared with the data from
  WMAP and ACBAR and to a standard $\La$CDM model (thin green
  line). The top panel uses a linear scale in $\ell$ while 
  the bottom panel used log scaling to emphasize the Sachs--Wolfe
  plateau at low values of $\ell$. The best fit parameter values used
for this plot are given in Table~\ref{t:results}.
}
\end{figure}

Even though the recent ACBAR data has not been used in the fitting
procedure, our best fit anisotropies shown in Fig.~\ref{f:pure1} do 
reproduce it nicely. 

\begin{figure}
\centerline{\epsfig{figure=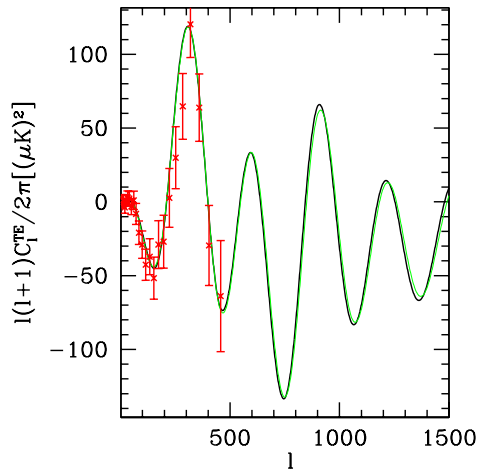,width=6.5cm}}
\caption{\label{f:pure2} The cross correlation spectrum of temperature
  anisotropy and E-polarization from a pure seed model is
  shown (fat black line) and compared with the WMAP data given in
  Ref.~\protect\cite{page} (we did not plot the data with $\ell>500$
  because of its large error bars). The best fit $\La$CDM curve
  is also indicated (fine green line) but is nearly invisible since
  it coincides nearly perfectly with the seed model curve. The
  parameter values are the same as for Fig.~\protect\ref{f:pure1}. }
\end{figure}

\begin{figure}
\centerline{\epsfig{figure=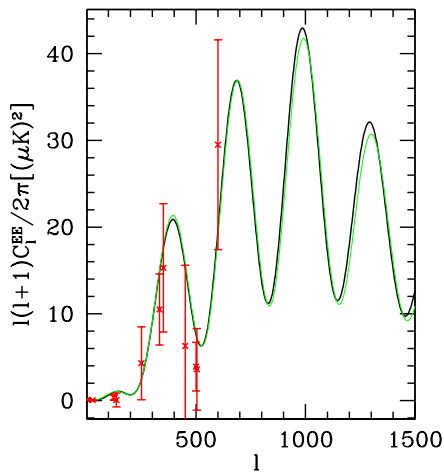,width=6.5cm}}
\caption{\label{f:pure3} The E-polarization from a pure seed model is
  shown (fat black line) and compared with the data from DASI,
  Boomerang-2003 and WMAP as given in
  Ref.~\protect\cite{page}. The best fit $\La$CDM polarization curve
  is also indicated (fine green line). The
  parameter values are the same as for Fig.~\protect\ref{f:pure1}. }
\end{figure}

In Fig.~\ref{f:pure2} we show the T-E-polarization cross
correlation for this model and compare it with the data and with the
result for a standard $\La$CDM model. As one already sees by eye,
within the accuracy of present data both models fit equally
well. The same is true for the E-polarization spectrum shown in
Fig.~\ref{f:pure3}. Since the perturbations are purely scalar there is no
B-polarization. 

\begin{figure}
\centerline{\epsfig{figure=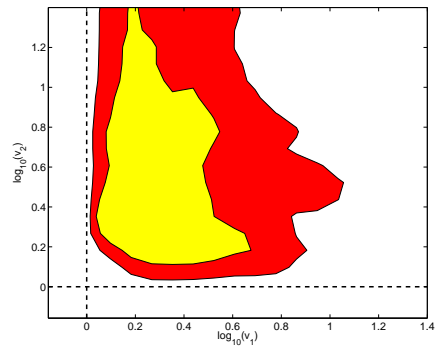,width=6.5cm}}
\caption{\label{f:v1v2} The 2-parameter likelihood plot for 
  ($v_1,v_2$) is shown (68\% and 95\% confidence contours).
Both velocities have to be larger than $1$. $v_1$ has a
preference for $v_1\approx 1.6$ while $v_2$ has no strong
upper limit.}
\end{figure}

It is interesting to note that the first polarization peak at $\ell
\simeq 130$ is also reproduced by the seed model. According
to~\cite{SpeZa} this is only possible since the explosions are
super-luminal, $v_{1,2} >1$. This is exactly what we see. As long as both 
velocities are below the speed of light, $v_1,v_2\le 1$, the first 
polarization peak remains absent, see Fig.~\ref{f:EEcausal}. When the 
velocities exceed the speed of light, the peak starts building up.
In order to match the observed T-E anti-correlation, which is in inflationary
models due to a superposition of a cosine wave (from the adiabatic density
mode) and a sine wave (from the velocity mode) and appears for
$k t_{\rm dec}\approx 0.66$ we need a velocity $v\gsim1/(kt_{\rm dec})\approx 1.5$
at decoupling, which agrees well with the point at which the
expanding shell model becomes acceptable. It is also interesting to note 
that the spectra do not depend on $v_2$ any more once it exceeds
about $v_2\sim 6$. This can be seen in the likelihood plot 
Fig.~\ref{f:v1v2}.
The likelihood plots for cosmological parameters are quite similar to the 
ones from inflation. For completeness we show some of them in the appendix.

In Table~\ref{t:results} we summarize the results for the acausal 
expanding shells model. The best-fit likelihood is slightly below
the one of the best-fit inflationary model with $\Delta\ln{\cal L}=2.3$.
Note that only about 18\% of the inflationary models have a $\Delta\ln{\cal L}$
of less than $2.3$, roughly agreeing with expectations for a $\chi^2$ distribution 
with $6$ degrees of freedom. We expect that the likelihood could 
be further improved, at the expense of introducing
more parameters, e.g. by allowing for a different evolution in matter
and radiation domination beyond the simple factor $1/(\HH t)$ in 
Eq.~(\ref{eq:frx}), or by allowing the shell velocities to change with time.
\begin{table}[h!] 
\begin{center}
\begin{tabular}{|c|c|c|c|c|c|}
\hline
\hline
 $v_1$   & $v_2$& $10\Om_mh^2$  &  $10\Om_bh^2$ & $H_0$ & $\tau$  \\
\hline
$1.65^{+7.1}_{-0.35}$ &$5.66^{+\infty}_{-4.26}$ &$1.34^{+0.07}_{-0.08}$ &
 $0.23^{+0.01}_{-0.01}$ &
$75^{+3}_{-3}$ &$0.11^{+0.07}_{-0.04}$ \\
\hline
\hline
\end{tabular}
\caption{Best-fit values and 95\% symmetric confidence intervals for the
acausally expanding shell model.
The best fit likelihood is $\ln{\cal L}= -1750.4$; slightly worse than the 
for simple inflationary models with the same number of parameters where we find
$\ln{\cal L} = -1748.1$.  \label{t:results}}
\end{center}
\end{table}

\begin{figure}
\centerline{\epsfig{figure=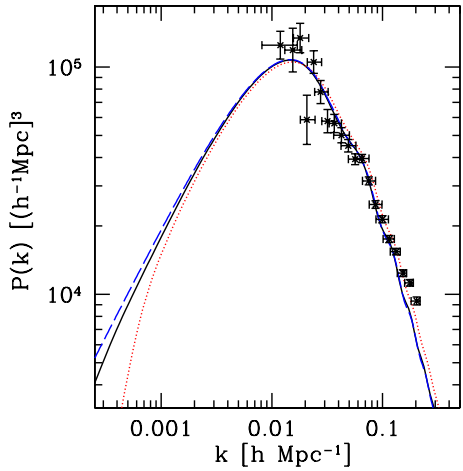,width=6.5cm}}
\caption{\label{f:pk}  The power spectra from causal expanding 
shells (dotted, red), acausal expanding shells (solid, black) and inflation 
(dashed, blue) are compared with data from luminous red 
galaxies~\protect\cite{lrg}. In the observable region, the acausal shell 
and inflationary power spectra are indistinguishable.}
\end{figure}

In our MCMC we have also fitted the power spectrum of luminous red
galaxies (LRG) as given in~\cite{lrg}. The best fit power spectra
are compared with the data in Fig.~\ref{f:pk}, for the causal and acausal shell 
models as well as for inflation. On super-horizon scales, 
$k\lsim 10^{-3}h$Mpc$^{-1}$, the power 
spectrum of the causal model is severely suppressed. Of course,
there is no data available on these scales.

\vspace{1cm}

 If we  require causality,  $v_1,v_2 \le 1$, we cannot fit the
CMB data with a pure seed model. However a mixture of expanding shells and
inflation can provide a good fit. 
Due to problems in the MCMC for the hybrid model, we do not have much statistics, so not all best fit parameters from the likelihood-analysis are converged values. 
The best fit parameters which we found for a
hybrid model with flat spatial sections are
\begin{eqnarray*}
10^{10}\ep^2 = 1.20\,, & v_1 = 0.80\,, & v_2 = 0.77\,, \\
10^{10}A_s = 20.02\,, &  n_s= 0.95\,, & \\
\Om_mh^2=  0.131\,, & \Om_bh^2=  0.0217\,, &  h=0.70 \,.
\end{eqnarray*}

\begin{figure}
\centerline{\epsfig{figure=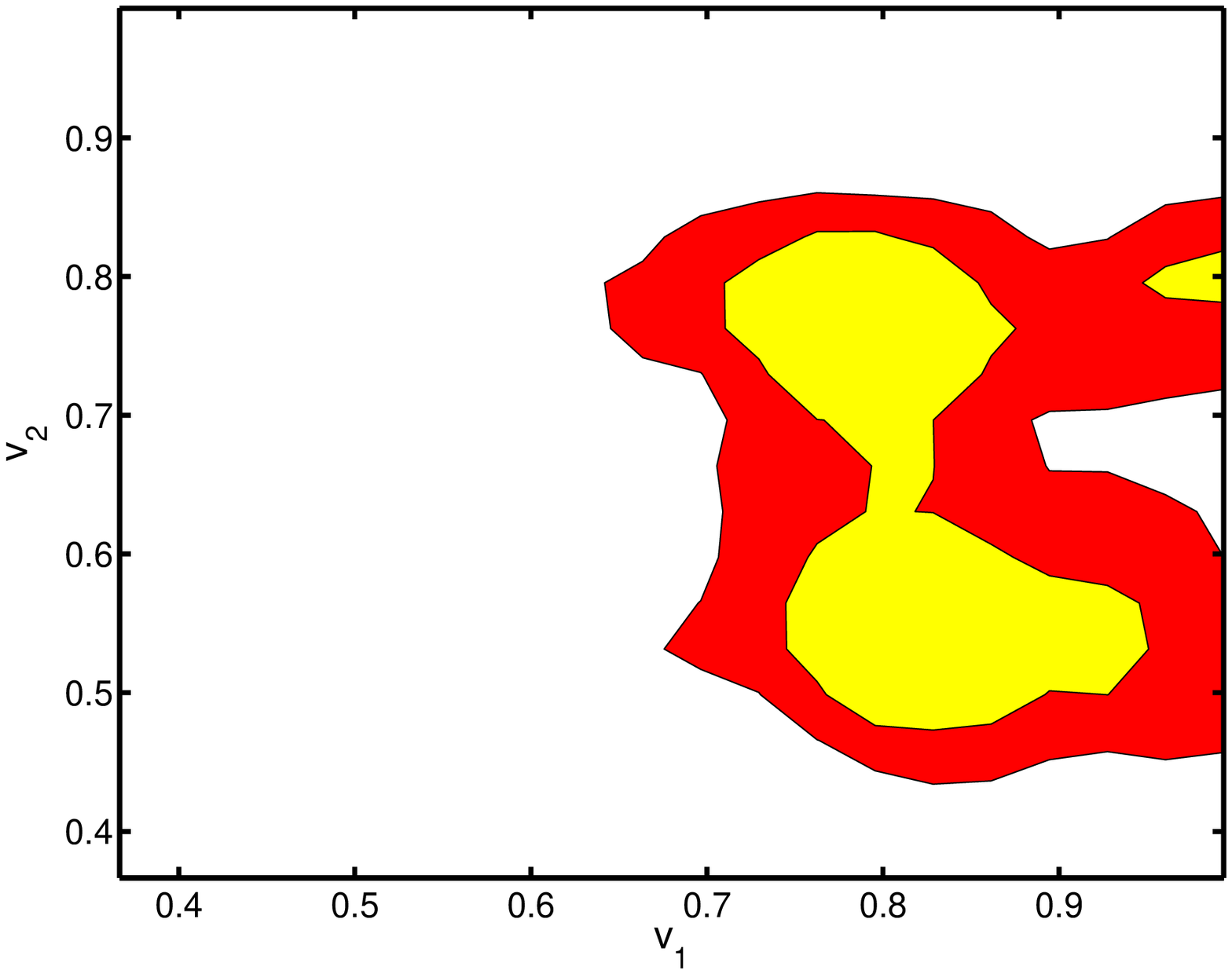,width=6.5cm}} 
\centerline{\epsfig{figure=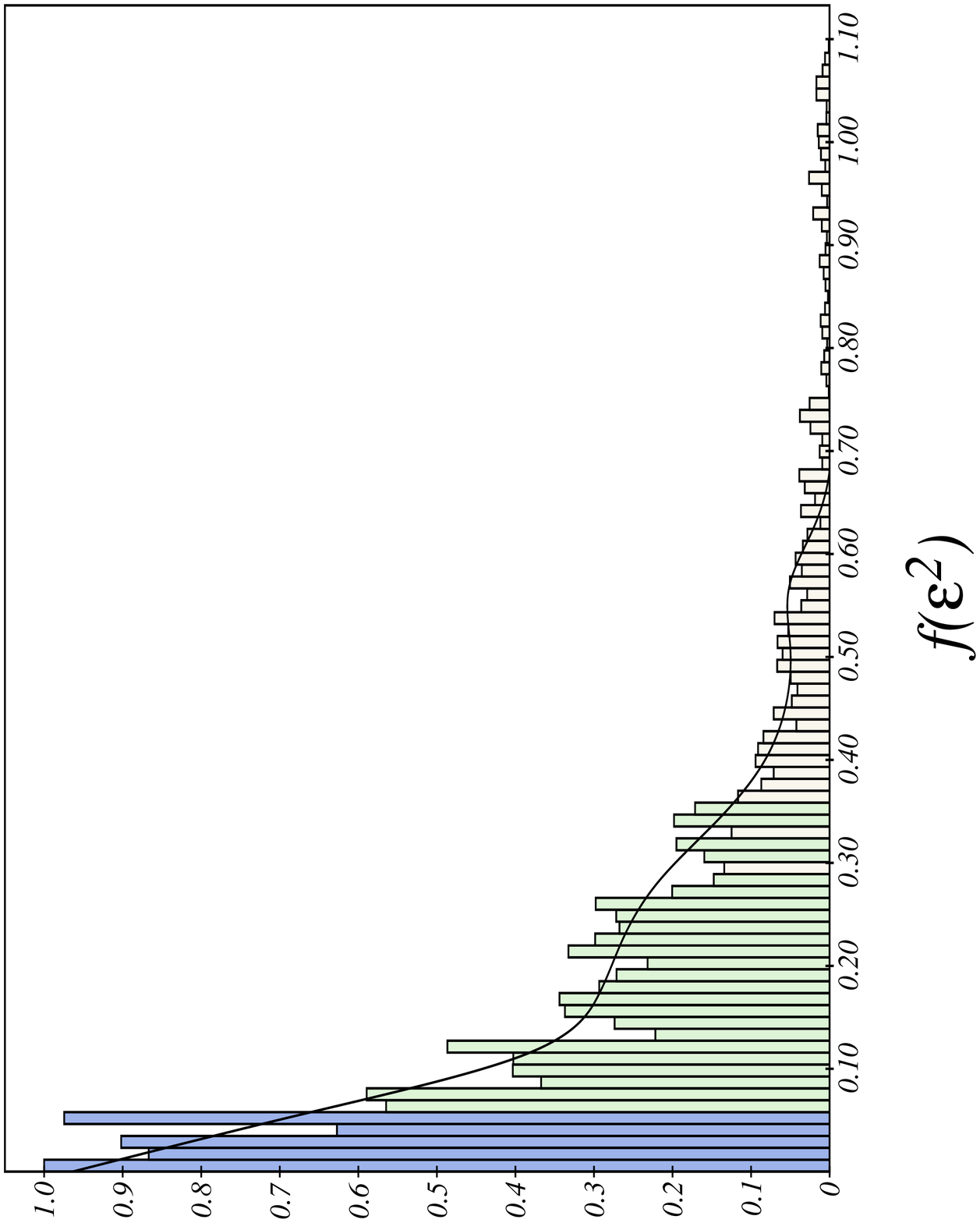,width=5.5cm,angle=-90}}
\caption{\label{f:v1v2hybrid} The 2-parameter likelihood plot for 
  ($v_1,v_2$) is shown for the mixed model with expanding shells and 
inflation (top). In the bottom panel we also show the likelihood 
distribution for $f\left(\ep^2\right)=\ln\ep^2-2\tau$.}
\end{figure}

\begin{figure}
 \vspace{1cm}
\centerline{\epsfig{figure=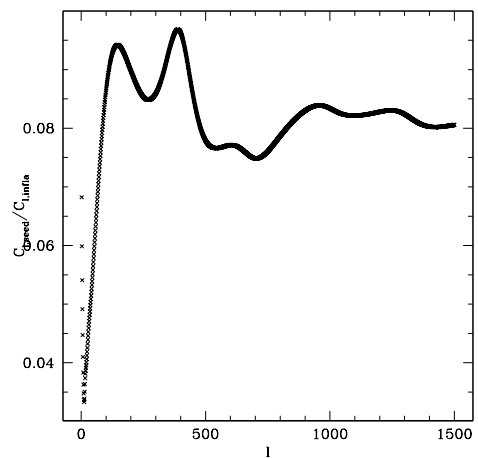,width=6.5cm}}
\caption{\label{f:ratio} The ratio 
$\frac{C^{TT}_{\ell,\mr{seed}}}{C^{TT}_{\ell,\mr{infla}}}$ 
for the best-fit hybrid model.}
\end{figure}

In Fig.~\ref{f:v1v2hybrid} we show the two parameter likelihood for the
shell velocities and the one parameter likelihood for the amplitude
$\ep^2$.
The likelihood for $\ep^2$ peaks close to zero, which seems to indicate that
the data prefers a vanishing contribution from the shells. Surprisingly for 
such a small amplitude (compared to the amplitude $A_s$ of the inflationary 
part) the contribution of the shells is non-negligible, this can be seen by 
looking at the ratio of the $C_\ell$ from shells and from inflation; see 
Fig.~\ref{f:ratio}.

As can also be seen for the best-fit causal model in Fig.~\ref{f:TTcausal}, 
the shells 
have a minimal contribution to the temperature anisotropies at 
$\ell\approx10$ so they do not contribute much at the Sachs-Wolfe plateau, 
but their contribution amounts to $\sim 10\%$ at higher $\ell$, as
shown in Fig.~\ref{f:ratio}. 
  
In all, this mixed model is is similar to models mixing inflation with 
topological defects. It is a logical possibility, but does not seem
very attractive since it increases the number of parameters with only
a marginal enhancement of the  likelihood of the model. 

\section{Conclusions}
\label{conc}
In this work we have revisited models with seeds, comparing them with
recent CMB data. We have specifically analyzed a model proposed by Neil
Turok~\cite{turok}, where the seeds are rapidly expanding spherical 
shells. We
have found that a seed model with sub-luminal velocities cannot fit
the CMB data. 

However, if we allow for super-luminal explosion
speeds, acausal shells, we can find an excellent fit to CMB anisotropies 
and polarization as
well as to the linear matter power spectrum. It is intriguing that the
velocities do not need to be {\em much} larger than the speed of light,
just $v\gsim1.5c$ is already sufficient. This model has effectively
the same number of parameters (two) as the simplest inflationary model
with purely scalar perturbations. The power spectra are so similar to 
the inflationary ones that it is not clear, at least on the level of
linear perturbations, how this model could be distinguished from 
inflationary perturbations. One possibility might be via a tensor 
component. Even though slight deviations from spherical symmetry might 
also lead to a tensor component for the expanding shell model, this 
component will probably not have the same characteristics as a tensor 
component from slow roll inflation (e.g. the consistency
relation between the tensor to scalar ratio and the tensor spectral index).

Super-luminal explosions do seem somewhat unphysical. Nevertheless,
it has been argued~\cite{MuBa} that super-luminal speeds in cosmology
do not lead to serious acausalities since Lorentz invariance is
broken in a Friedmann-Lema\^\i tre universe, where the cosmological 
reference frame represents a preferred frame. Although this seems 
quite artificial on small scales, the argument may be valid on the 
cosmologically large scales of these expanding shells. It might therefore be
advisable to keep an open mind, especially when considering that inflation 
is usually implemented with the help of the potential energy of a scalar field,
the normalization of which is intimately linked to the cosmological 
constant, the probably biggest unsolved problem in cosmology.

We have also investigated hybrid models with seeds and inflationary
perturbations. 
Good fits are obtained with seed contributions of about 10\% on
angular scales, $\ell \gsim 100$.

\begin{acknowledgments}
This work is supported by the Swiss National Science Foundation. Ruth
and Sandro thank Sussex University for hospitality. The numerical
computations have been performed on the Myrinet cluster of Geneva
University and Archi cluster of the University of Sussex. 
\end{acknowledgments}

\appendix
\section{ Likelihoods}
\begin{figure}
\centerline{\epsfig{figure=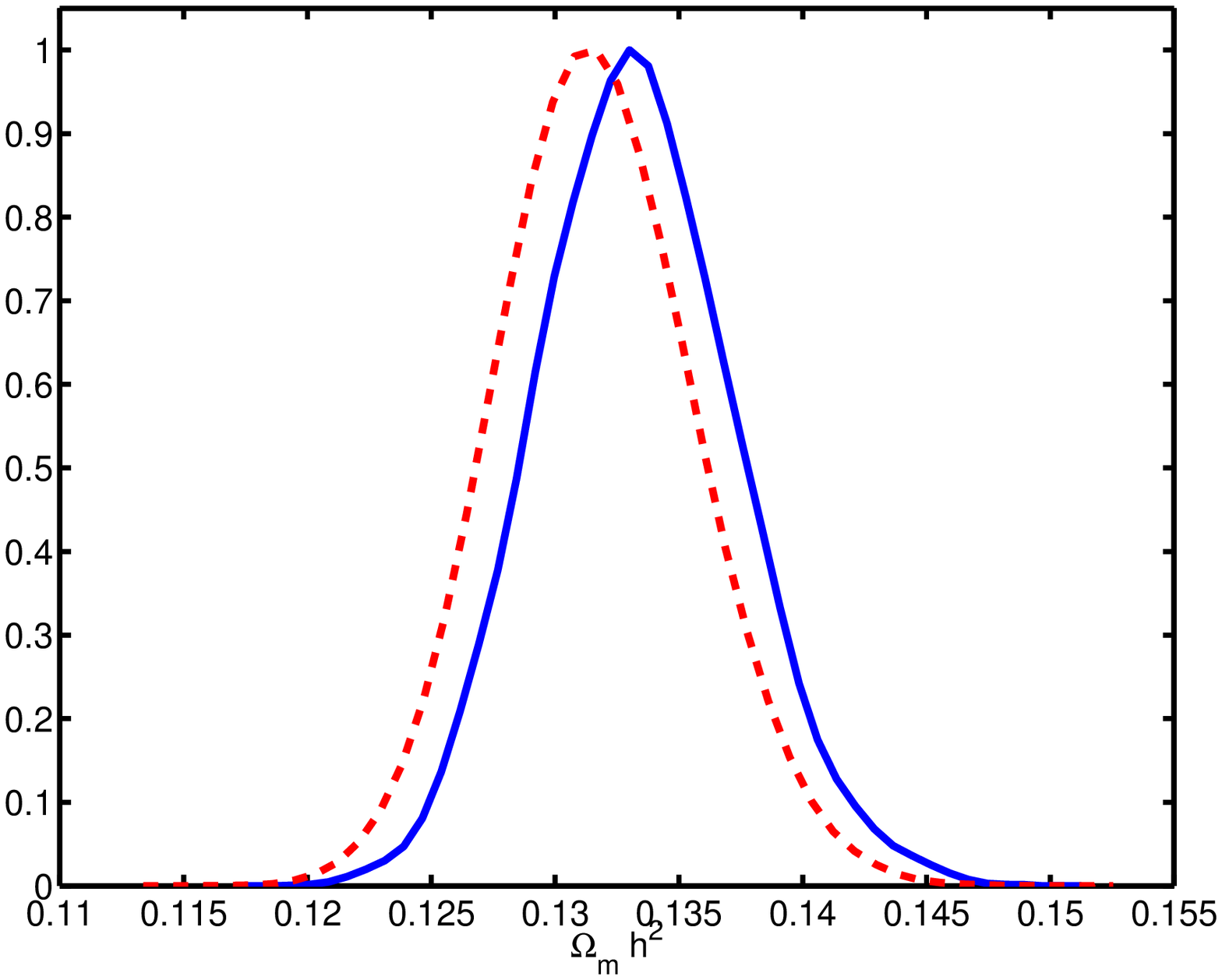,width=7cm}}
\centerline{\epsfig{figure=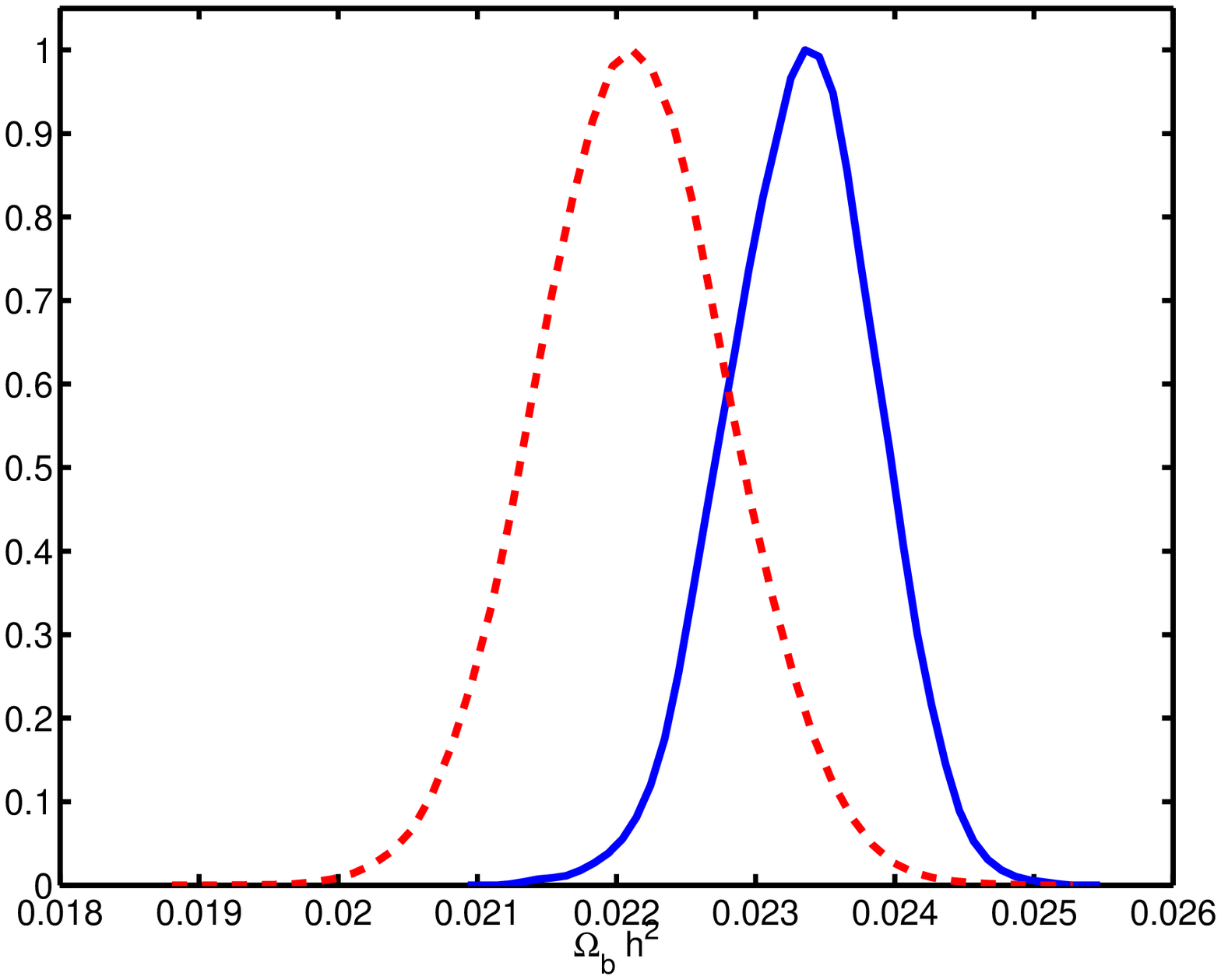,width=7cm}}
\centerline{\epsfig{figure=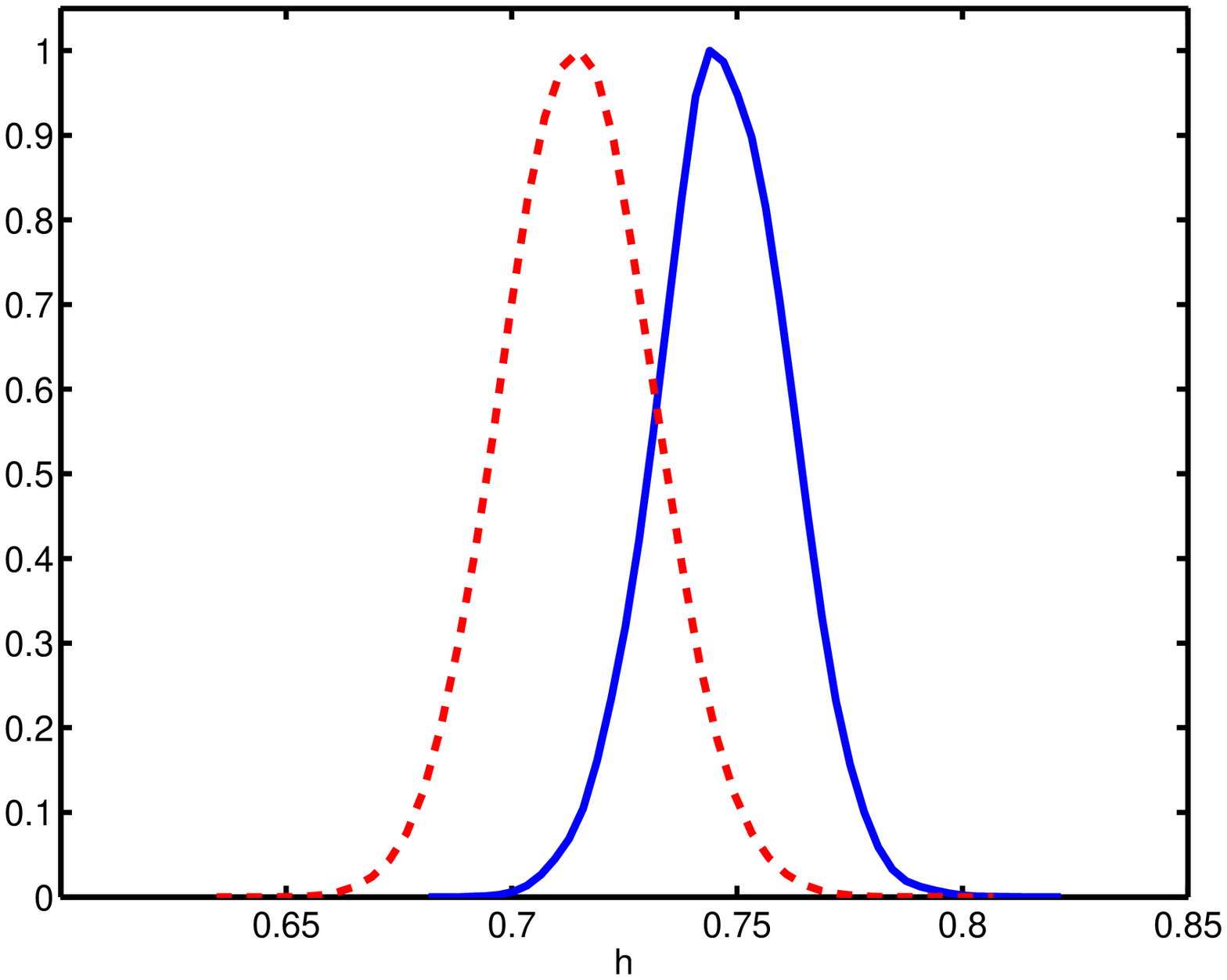,width=7cm}}
\centerline{\epsfig{figure=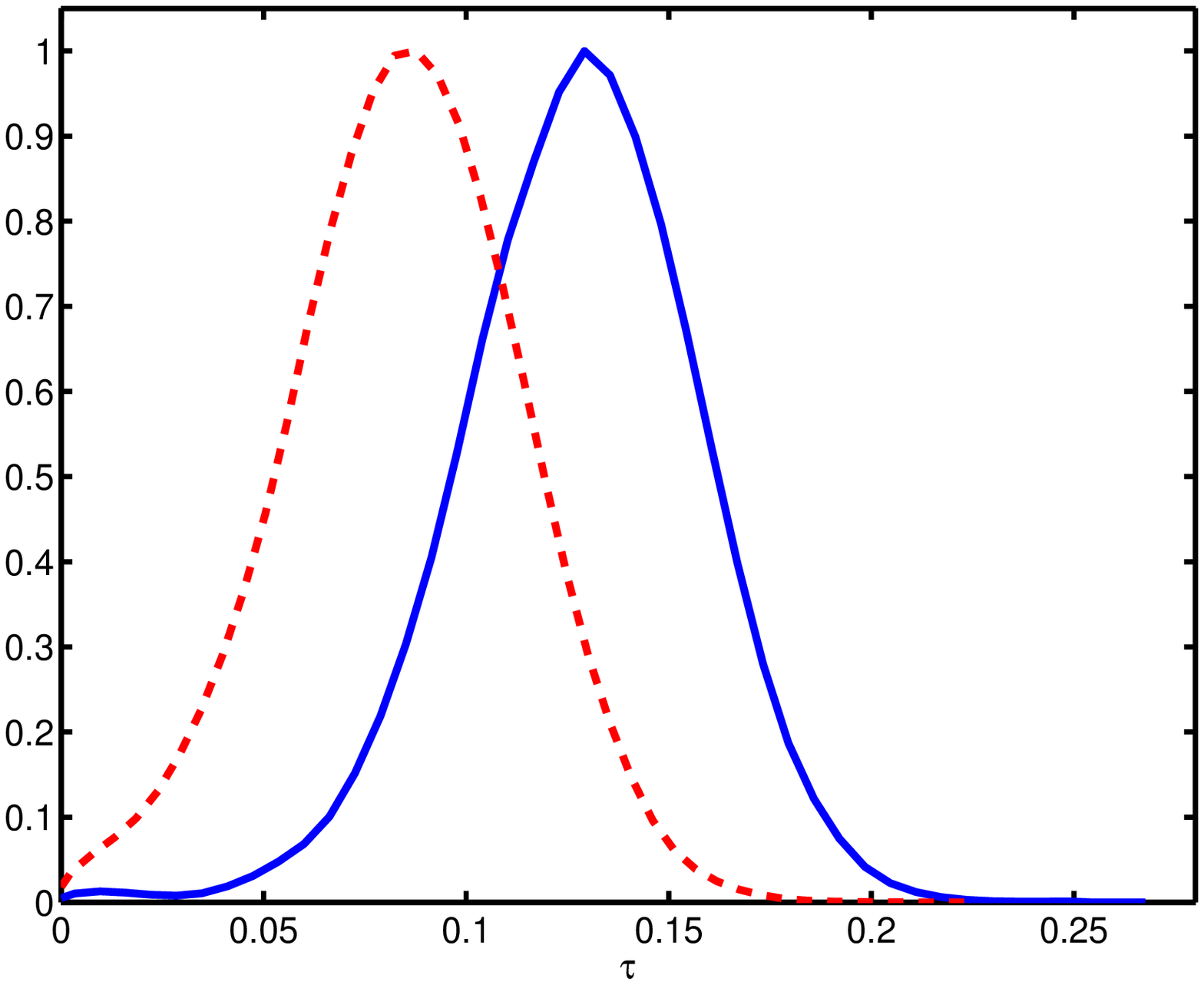,width=7cm}}
\caption{\label{f:cospar} The 1D likelihood plots for the acausal model 
(blue solid line) and the standard inflationary
case (red dashed line). The top-left panel shows $\Omega_m h^2$, the
top right panel $\Omega_b h^2$, the lower left panel $h$ and the
lower right panel $\tau$.}
\end{figure}

In this appendix we show some additional likelihood
plots for the parameters of the acausal expanding shell model discussed 
in this work. For the 1D likelihoods, the (blue) solid line always shows
the marginalized likelihood of the acausal model while the (red) dashed
curve is the marginalized likelihood for the inflationary case.  In 
Fig.~\ref{f:cospar}  one sees that the cosmological parameters obtained 
for the acausal seed model are quit similar to inflationary parameters.
Even though the best fit $\Om_bh^2$, $h$ and $\tau$ for the seed model
are somewhat higher, the inflationary best fit value is within one sigma.

\begin{figure}
\centerline{\epsfig{figure=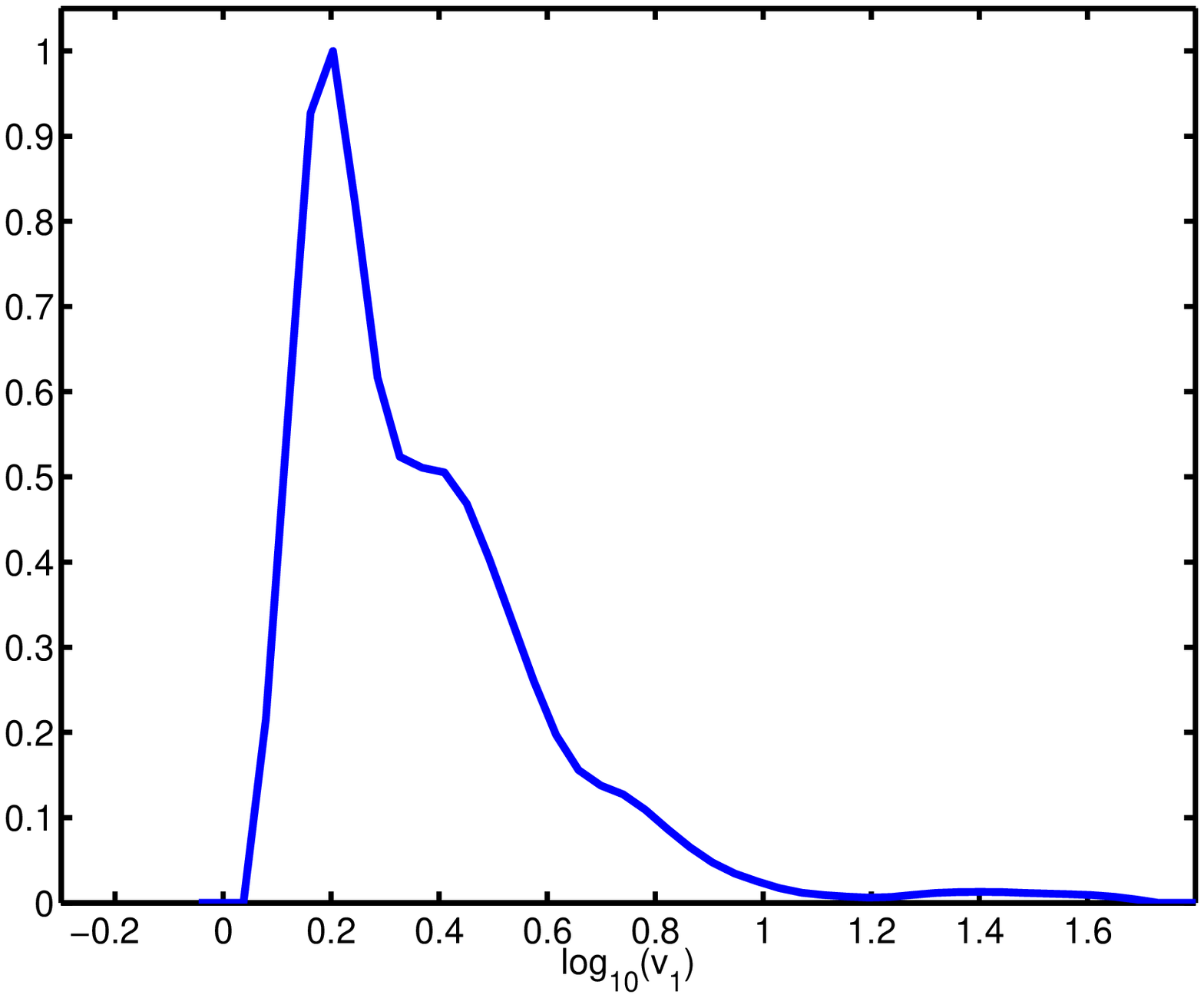,width=7cm}}
\centerline{\epsfig{figure=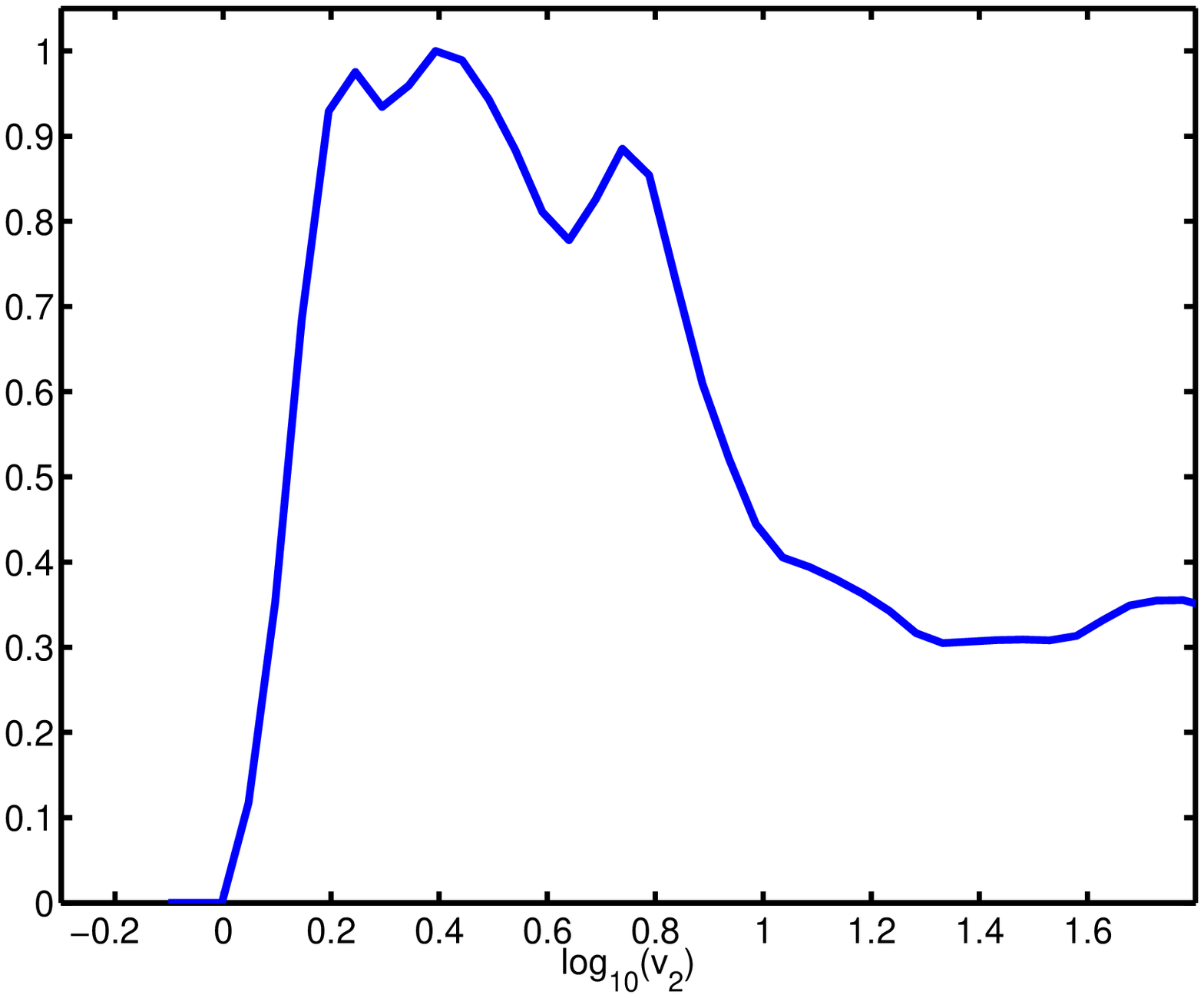,width=7cm}}
\caption{\label{f:velpar} The 1D likelihood plots for the velocities
in the acausal model: The left panel shows $\log_{10}(v_1)$ and
the right panel $\log_{10}(v_2)$. $v_1$ shows a preference for 
$v_1\approx 1.5$ with a long tail to higher velocities, while 
$v_2$ is basically unconstrained apart from $v_2>1$.}
\end{figure}

In Fig.~\ref{f:velpar}  the 1 dimensional likelihoods for the velocities
are shown. Clearly, once the velocities are above about 1.5, the fit
becomes good. For $v_1$ the likelihood decreases steeply above 
about $v_1=3$ (albeit with a long tail)  
while it remains nearly constant for $v_2$ which therefore 
seems not to have an upper bound. We have also found that above
$v_2\sim 6$  the spectra are nearly independent of the value of $v_2$.

\end{document}